# Electric Field-Dependent Charge-Carrier Velocity in Semiconducting Carbon Nanotubes

Yung-Fu Chen and M. S. Fuhrer

Department of Physics and Center for Superconductivity Research, University of Maryland, College Park, MD 20742-4111, USA

Charge transport in semiconducting single-walled nanotubes (SWNTs) with Schottky-barrier contacts has been studied at high bias. We observe nearly symmetric ambipolar transport with electron and hole currents significantly exceeding 25 µA, the reported current limit in metallic SWNTs due to optical phonon emission. Four simple models for the field-dependent velocity (ballistic, current saturation, velocity saturation, and constant mobility) are studied in the unipolar regime; the high-bias behavior is best explained by a velocity saturation model with a saturation velocity of $2 \times 10^7$ cm/s.

One of the most striking aspects of carbon nanotubes is that they can carry extremely large current densities [1-3], exceeding $10^9$ A/cm$^2$, orders of magnitude larger than those at which metal wires fail by electromigration [4]. Zone-boundary phonon emission at energy $hf = 160$ meV explains the current limit in metallic single-wall carbon nanotubes (SWNTs) of $(4e/h)(hf) \approx 25$ µA[1,5,6]. By using electrical breakdown of multi-wall carbon nanotubes (MWNTs), it was found that metallic and semiconducting shells carry similar saturation currents [2,7]. However, there is no reason to assume that the limit in semiconducting nanotubes is identical, since the electronic band structure is different: The equivalent zero-momentum phonon emission process in semiconducting nanotubes involves relaxation of electrons across the band gap. Bourlon, *et al.* proposed a model of competition between electron-phonon scattering and Zener tunneling to explain the geometrical dependence of saturation currents in metallic and semiconducting single shells in MWNTs [8], but experiments on single semiconducting SWNTs are lacking.

In this letter we investigate high-bias transport in single Schottky-barrier (SB)-contacted semiconducting SWNTs up to bias voltages of 10 V, corresponding to average electric fields up to 5 kV/cm. We examine four simple models for the electric-field dependence of the carrier velocity: ballistic, current saturation, velocity saturation, and constant mobility to understand the current at unipolar bias voltage regimes. The results are directly compared with the experimental data, and we find the best agreement with the velocity saturation model, with a saturation velocity $v_s$ of $2 \times 10^7$ cm/s.

Heavily n-doped Si chips with a SiO$_2$ layer of thickness $t = 500$ nm are used as our device substrates. SWNTs were grown using a chemical vapor deposition (CVD) process with catalyst prepared as in Ref. 9 and carbon feedstock gases as introduced as in Ref. 10. A scanning electron microscope (SEM) was used to locate the nanotubes [11], and electron-beam lithography followed by thermal evaporation of 1.5 nm Cr and 100 nm Au formed the electrodes. Devices were annealed at 400 °C under Ar and $H_2$ flow to lower the contact resistance. The diameters of the nanotubes were checked by atomic force microscope (AFM).

We have investigated several semiconducting SWNT devices with similar transport behavior [12]. The measured nanotube diameters $d$ range from 2 to 2.4 nm, and the nanotube lengths between the contacts (channel length $L$) range from 10 to 20 µm. The measured nanotube diameters indicate they are SWNTs or small MWNTs. (We expect that in the case of a small MWNT with all semiconducting shells that the transport occurs primarily in the outermost shell with the lowest bandgap.) We focus on the transport data from one particular device, with $d$ = 2.4 nm and $L$ = 20 µm.

Electrical measurements were performed by grounding the source electrode ($V_s$ = 0) and applying $V_d$ to the drain, and $V_g$ to the gate, while measuring the drain current $I_d$. Figure 1 shows $I_d$ as a function of $V_g$ at $V_d$ = -1 V. This device behaves as an ambipolar semiconductor although p-type (hole) conduction is slightly better than n-type (electron) conduction. Because of the ambipolar behavior, we assume the presence of SBs at the electrodes for electrons and holes, but a smaller SB for holes. The maximum transconductances $g_m$ for holes and electrons are ~ 2.5 and 2.0 µA/V, respectively, which correspond to field-effect mobilities [13] $\mu_{FE} \approx 2.5 \times 10^4$ and $2.0 \times 10^4$ cm$^2$/Vs (here we have used the electrostatic gate capacitance per length $c_{g,e} \approx 2\pi\varepsilon\varepsilon_0/\ln(4t/d) \approx 0.2$ pF/cm where $\varepsilon_0$ is the electric constant, $\varepsilon \approx 2.45$ the average dielectric constant of the oxide and vacuum). The on-state conductances $G \approx 13$ and 7 µS for hole and electron indicate mean-free-paths $l = LG/2G_0$ of at least 1.6 and 0.9 µm. These values are comparable to the highest measured values for SWNTs [13] though contact resistance may play a significant role in this case.

Figure 2 shows measured $I_d$ vs. $V_d$ up to ±10 V at different gate biases $V_g$. We note several striking features of the data. First, the data is highly symmetric under reversal of both $V_d$ and $V_g$, indicating ambipolar behavior. Second, the electron and hole currents significantly exceed 25 µA with no obvious evidence of saturation.

At finite positive (negative) gate voltages, the curves of Figure 2 show typical n-type (p-type) transistor behavior, i.e. saturation of $I_d$ at positive (negative) drain voltage. The saturation is followed by an increase in the current as $V_d$ becomes greater than $2V_g$. We interpret this increase in $I_d$ as the change from majority electron to majority hole current or vice versa. $V_d = 2V_g$ corresponds to the symmetric bias condition [14,15], equivalent to holding $V_g = 0$ and applying equal and opposite voltages to source and drain; at this point electron and hole currents should be equal. The inset of Figure 2 shows the current is minimum at $V_d = 2V_g$. The curves for electrons and holes are similar, with slightly better conduction for holes, consistent with the low-bias data of Figure 1.

We will focus here on the conduction at unipolar bias regions ($V_g$ is larger than both $V_d$ and $V_s$ for electrons, or $V_g$ is smaller than both $V_d$ and $V_s$ for holes). Under these bias conditions, there are very few minority carriers in the channel. (The ambipolar case, where the current contribution from minority carriers can not be ignored, will be discussed elsewhere.)

First we examine the expected behavior for a ballistic nanotube FET. We calculate the current using some assumptions and simplifications: (1) The nanotube is undoped. (2) The electrode Fermi level is aligned with the middle of the nanotube gap. (3) The subbands are approximated by a hyperbolic band structure $E(k) - (-e)V_{NT}(x) = \mp\sqrt{(\hbar v_{F,m} k)^2 + (nD)^2}$, where the upper (lower) sign is for the valence (conduction) bands, $V_{NT}(x)$ is the potential of the nanotube at position $x$, $k$ is the wave vector, $v_{F,m} = 9.35 \times 10^5$ m/s is the Fermi velocity of metallic nanotubes, $n = 1$ for first subbands, $n = 2$ for second subbands, and $D$ is a half of nanotube bandgap. (4) The device capacitance is dominated by capacitance to the gate since the channel length is fairly long compared with gate oxide thickness, and the charge density $q(x) = \pm n(x)e = c_{g,e}(V_{NT}(x) - V_g)$ is determined locally, where $n(x) = 4\int dk(f(E(k_>) - \mu_s) + f(E(k_<) - \mu_d))$ is the carrier density (the 4 from spin and subband degeneracies), $k_>$ ($k_<$) means $k$ is in the range of $k > 0$ ($k < 0$). (5) Zero temperature. (6) Contact resistance is neglected; i.e. the SBs at the contacts are transparent once the bias across the SBs exceeds one-half the band gap. The last two assumptions are reasonable for describing transport at biases greatly exceeding the SB height and the temperature. $V_{NT}$ and $n$ at the contacts are determined by satisfying the assumption (3), (4), (6) and bias conditions self-consistently. The current is then calculated by $I = \mp 4e\int dk(f(E(k_>) - \mu_s) + f(E(k_<) - \mu_d))v(k)$, where $v(k) = \frac{1}{\hbar}\frac{\partial E(k)}{\partial k}$ is the velocity of state $k$. Here, only the lowest conduction bands or highest valence bands are considered to contribute the current conduction.

The result of the ballistic model for holes as majority carriers is plotted in Figure 3(b). At low drain bias, the conductance is $\frac{4e^2}{h} = 155$ µS, which is the signature of a ballistic SWNT. At high positive drain bias, the current increases more slowly, with near-constant slope $\frac{c_{g,e}}{c_{g,e} + c_q} \frac{4e^2}{h} \approx 17$ µS, where $c_q = e^2 D(E)$ is the quantum capacitance, with $D(E)$ the energy-dependent density of states. Except for Fermi energies very near the band edge [16] $c_q$ is well approximated by the quantum capacitance of a metallic SWNT $c_{q,m} = 8e^2/hv_{F,m} \approx 3.31$ pF/cm. In the ballistic case, only the right-moving branch of the conduction band is filled, and $c_q = c_{q,m}/2 \approx 1.65$ pF/cm. Here the hole density is limited by the voltage difference between gate and drain. At negative high drain bias, the current saturates because no more holes are added by increasing negative drain bias, since now the amount of holes is controlled by the voltage difference between gate and source.

The ballistic model reproduces many of qualitative features of the experimental data in Figure 3(a), but the current and conductance are significantly higher than those in the experiment. This is reasonable: since the channel length is tens of microns, we do not expect ballistic transport. We next examine two models for the field-dependent velocity in the nanotube. In the first, the current saturation model, we assume that the maximum difference in the left- and right-moving quasi-Fermi levels is set by optical phonon scattering at $hf \approx 160$ meV. The current is then limited to $4ef \approx 25$ µA as observed in metallic nanotubes [1], and as has been suggested for semiconducting nanotubes [17]. The empirical I-V relation $\frac{V}{I} = R_0 + \frac{|V|}{I_0}$, where $I_0 = 25$ µA is the saturation current for metallic SWNT, was suggested in Ref. 1. If we represent this relation in terms of mobility $\mu$ and electric field $E$, we have:

$$\mu^{-1} = \mu_0^{-1} + ne\frac{|E|}{I_0}, \tag{1}$$

where $\mu_0$ is the zero field mobility and $n$ is the carrier density.

Again using the assumptions above, we calculate the current $I = q(x)\mu E(x)$, where $E(x) = -\nabla V_{NT}(x)$ and $q(x) = c_{g,e}(V_{NT}(x) - V_g)$ are functions of position $x$. The result of the current saturation model for hole conduction is shown in Figure 3(c), where the fitting parameters are $I_0 = 25$ µA and zero field hole mobility $\mu_0 = 2.7 \times 10^4$ cm$^2$/Vs. (The zero-field mobility $\mu_0$, determined using the data in the range -0.1 V < $V_d$ < 0.1 V, is slightly larger than the field effect mobility $\mu_{FE}$ discussed above due to finite $V_d$ = -1 V used there.) Figure 3(a) compares the calculation with experiment; disagreement is seen in several aspects. First, rather than saturating at 25 µA at high positive drain bias, the measured current increases with roughly constant slope. Second, the current at negative bias is larger than that in experiment. $I_0 = 25$ µA was chosen assuming only the first valence subbands participate in the conduction and the physics is similar to metallic SWNT; however, no choice of $I_0$ gives a good fit. Specifically, $I_0 = 50$ µA, which might correspond to two contributing subbands, is significantly worse.

The qualitative agreement of the ballistic model with the measured current suggests that the charge-controlled model of the nanotube transistor, namely $I_d \propto q$, is correct. However, the charges must not move with Fermi velocity, but somewhat slower. In conventional semiconductors, typically the electric-field-dependent carrier velocity is observed to saturate to a constant value at high electric field. Empirically, the carrier mobility often follows

$$\mu^{-1} = \mu_0^{-1} + \frac{|E|}{v_s},\qquad(2)$$

where $v_s$ is the saturation velocity. Eq. (2) means at low E-field regime, the carrier velocity $v$ increases linearly with E-field with slope $\mu_0$; at high E-field regime, $v$ saturates at $v_s$. Perebeinos, *et al.* have calculated electron-phonon interaction within a tight binding model and derive mobility-E-field relation for a single electron [18], and Eq. (2) fits their result very well. However, it is not at all clear whether their results extrapolated to many electrons would give current saturation or velocity saturation (since these are identical in a one-electron model). The calculation of current under the velocity saturation model is the same as that in the ballistic model except $v = \mu E$, $\mu$ is described by Eq. (2), and the pinch-off effect is included [19].

The *I-V* curves of the velocity saturation model for hole conduction are plotted in Figure 3(d), where the fitting parameters are hole saturation velocity $v_s = 2 \times 10^7$ cm/s [20] and zero field hole mobility $\mu_0 = 2.7 \times 10^4$ cm$^2$/Vs. Simply speaking, $\mu_0$ is determined by fitting the current at low drain bias; $v_s$ is determined by fitting the current at high negative drain bias.

In order to compare the model, we plot in Figure 4 the saturation current $I_{sat}$ at different $V_g$ for the experimental data and the current saturation and velocity saturation models. Also included is a calculation carried out for a constant hole mobility of 5000 cm$^2$/Vs. The figure shows that the experimental behavior fits the velocity saturation model very well - the velocity saturation model predicts linear $I_{sat}$ vs. $V_g$, as seen in the data. The current-saturation model always produces sublinear $I_{sat}$ vs. $V_g$, while the constant mobility model has $I_{sat} \sim V_g^2$.

Though a simple model of velocity saturation describes the experimental data surprisingly well, some problems remain. First, contact effects are not considered, which causes poor agreement around $V_d = 0$ and $V_g = 0$. Second, the experimental current at high positive drain bias does not increase as fast as expected, indicating that $v_s$ may decrease slightly with increasing charge density. More study is needed to fully understand SWNT transistors under high-bias voltages.

The saturation velocity is smaller than the peak carrier velocity of $4.5 \times 10^7$ cm/s in Ref. 21 for a 2.4 nm diameter nanotube at an electric field of ~ 5 kV/cm, and $5 \times 10^7$ cm/s in Ref. 18 using a one-electron model, but is still more than twice as high as in silicon inversion layers [22].

The authors are grateful for support from the National Science Foundation under grant DMR-01-02950, and for useful discussions with Gary Pennington, Akin Akturk, and Neil Goldsman.

[19] Pinch-off happens because the carrier velocity is limited at $v_s$. To maintain constant current in nanotube, the channel needs finite amount of charge even when $V_d$ approaches $V_g$. We include the pinch-off effect in velocity saturation model as follows: when we calculate $I_d$ vs. negative $V_d$, the magnitude of current eventually starts to decrease as $V_d$ is beyond a threshold voltage $V_{d,th}$; i.e. the maximum of $I_d$, $I_{d,max}$, occurs at $V_{d,th}$. The current is fixed at $I_{d,max}$ when $V_d$ is greater than $V_{d,th}$. The same procedure is applied to the source when $V_d$ is positive.

[20] The unipolar majority-electron bias region ($V_g$ larger than both $V_d$ and $V_s$) is very similar to the majority-hole region, indicating symmetric conduction and valence band structure. By a similar analysis, we find the electron saturation velocity is also ~ $2 \times 10^7$ cm/s.

[21] G. Pennington and N. Goldsman, Physical Review B **68**, 045426 (2003).

[22] J. A. Cooper and D. F. Nelson, Journal of Applied Physics **54**, 1445 (1983).

Figure Captions:

Figure 1. Drain current $I_d$ as a function of gate voltage $V_g$ at drain voltage $V_d$ = -1 V and temperature 4.2 K. The solid and dashed lines correspond to linear (left) and logarithmic (right) scales, respectively.

Figure 2. Ambipolar semiconducting nanotube transistor. (a) Drain current $I_d$ as a function of drain voltage $V_d$ at gate voltages $V_g$ from -9 V to 9 V, in 1 V steps is measured. $V_d$ is applied up to ±10 V. Temperature is 4.2 K. The inset is a color-scale plot of $I_d$ as a function of $V_d$ and $V_g$, where red and blue colors represent positive and negative $I_d$, respectively. The dashed line in the inset indicates $V_g = V_d/2$; the current minimum occurs along this line.

Figure 3. Comparison between experimental data and simulation for several models with holes as majority carriers. (a) Experimental $I_d$ as a function of $V_d$ at $V_g$ from -9 V to -1 V, in 1 V steps. The thick curve is for $V_g$ = -9 V. (b) Ballistic model (c) current saturation model, and (d) velocity saturation model are plotted for the same gate voltages as the experimental data. Note the different vertical scale for (b).

Figure 4. Saturation current $I_{sat}$ at different $V_g$ for experimental data and theoretical models discussed in text. $I_{sat}$ is $I_d$ at $V_d = V_g$.

Figure 1

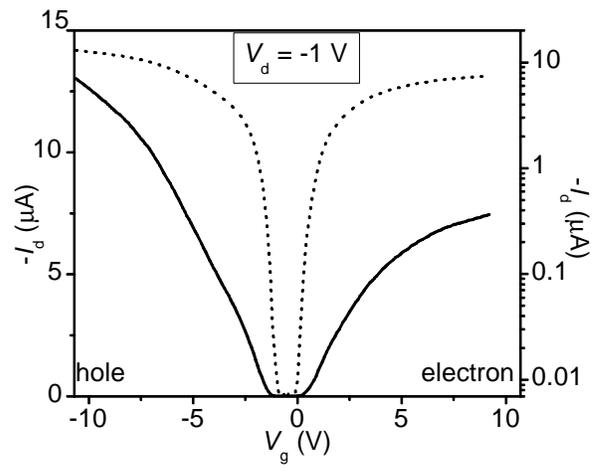

Figure 2

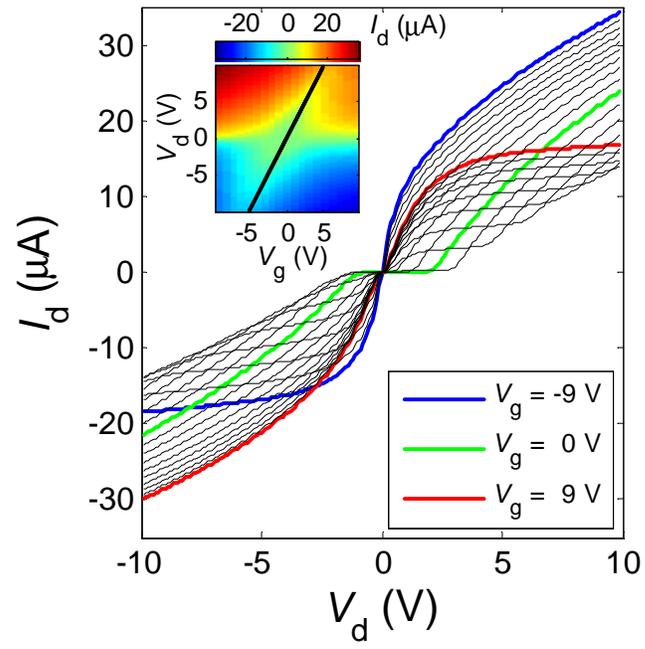

Figure 3

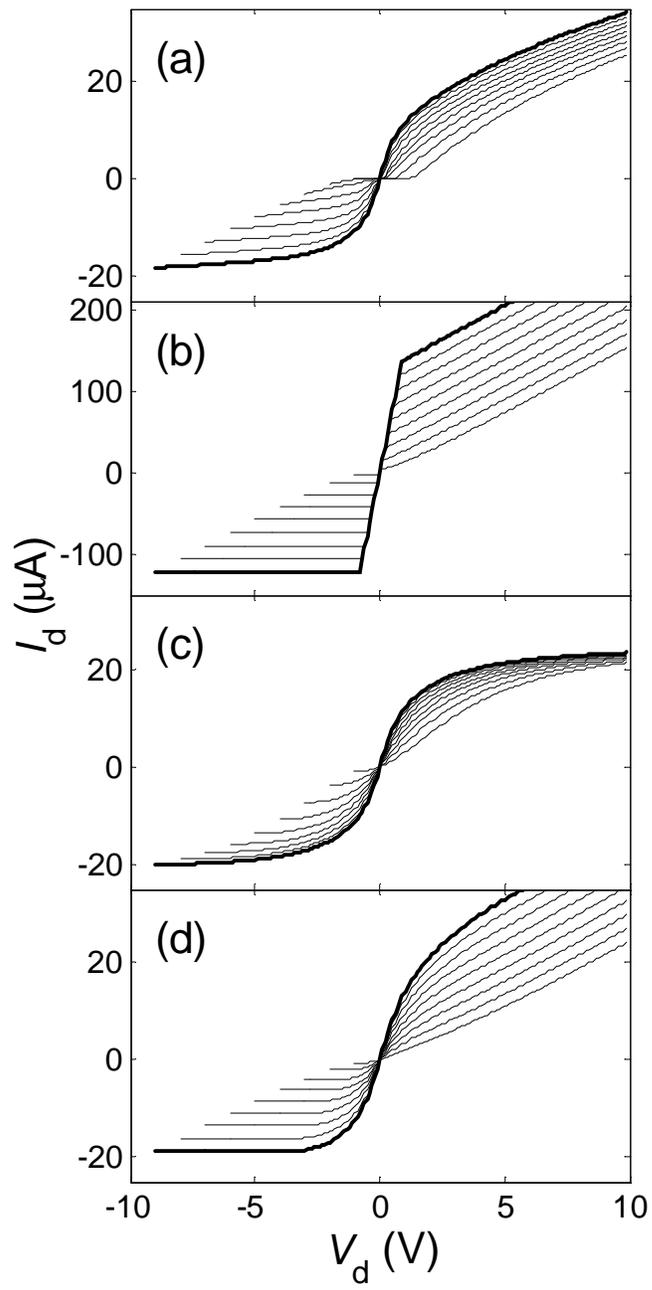

Figure 4

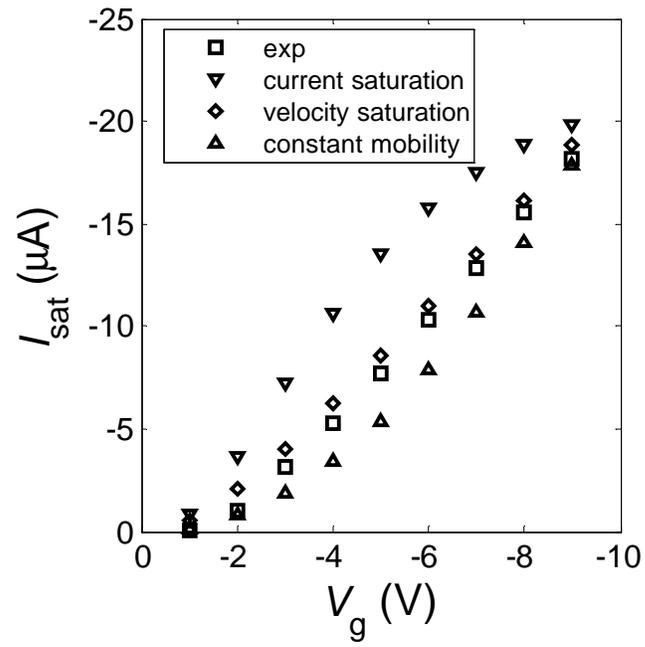